# Diagnosing Biases in Tropical Atlantic-Pacific Multi-Decadal Teleconnections Across CMIP6 and E3SM Models

**Yan Xia[1]\*, Yong-Fu Lin[1], Jin-Yi Yu[1], Walter Hannah[2], Mike Pritchard[1]**

[1]Department of Earth System Science, University of California, Irvine, California, USA

[2]Lawrence Livermore National Laboratory, Livermore, CA

Corresponding author: Yan Xia  (yxia24@uci.edu)

**Abstract**

Decadal-scale interactions between the tropical Atlantic and Pacific Oceans play a crucial role in global climate variability through bidirectional teleconnections. Current climate models show persistent biases in representing these basin interactions, particularly in simulating the Atlantic's influence on Pacific climate. Using historical simulations from 27 CMIP6 models and two configurations of the Energy Exascale Earth System Model (E3SM) during 1950-2015, we systematically evaluate tropical Atlantic-Pacific teleconnections through both Walker circulation and extratropical wave responses. Most models exhibit Pacific-dominated teleconnections, contradicting observational evidence of Atlantic control on Pacific variability during the past 40 years. By developing a performance metric that combines tropical circulation patterns and extratropical wave propagation, we identify two distinct model behaviors: high-skill models capture the bidirectional Atlantic-Pacific teleconnections with a secondary symptom of systematic 20-degree westward shifts in convective centers, while low-skill models display amplified Pacific dominance through reversed Walker circulation responses warming in both tropical basins. Comparative analysis between standard E3SMv2 and its multi-scale modeling framework configuration demonstrates that implementing more sophisticated cloud-scale processes alone, with limited model tuning, cannot resolve these teleconnection biases. Our results identify four CMIP6 models and E3SMv2 that effectively reproduce observed teleconnection pathways, offering a comprehensive diagnostic framework for evaluating decadal Atlantic-Pacific interactions in climate models.

# Introduction

Interactions between the Atlantic and Pacific oceans play a key role in global climate variability across multiple timescales. While current generation climate models effectively capture these interactions at interannual timescales, they misrepresent observed decadal scale patterns, particularly in simulating Atlantic influences on Pacific climate variability[1–4]. This model-observation discrepancy limits our ability to predict long-term climate trends and understand global warming patterns[1,5].

Recent syntheses argue that tropical Atlantic-Pacific interactions operate through two primary pathways on decadal timescales, as demonstrated by observational analyses[6]. In the Atlantic-to-Pacific pathway, Atlantic Multidecadal Variability (AMV) warming induces a La Niña-like pattern in the Central and Eastern Pacific (CEP) through coupled ocean-atmosphere processes[7]. The initial response manifests as an anomalous Walker circulation, with enhanced ascending motion over the Atlantic and subsidence over the CEP. The Atlantic warming also triggers an eastward-propagating Kelvin wave response, establishing a secondary convective center over the Indo-Western Pacific [5]. This convective center reinforces CEP subsidence through Walker circulation adjustments. These atmospheric responses strengthen easterly winds over the equatorial Pacific, which amplify the initial cooling through Bjerknes feedback[8]. In the Pacific-to-Atlantic pathway, positive Pacific Decadal Variability (PDV) generates an El Niño-like pattern in the CEP that modify Atlantic SSTs through both tropical and extratropical processes[6]. In the tropics, PDV-induced convective heating over the CEP modifies the Walker circulation, causing subsidence and surface wind changes over the Atlantic. In the extratropics, PDV excites Pacific-North American (PNA) and Pacific-South American (PSA) wave trains, which induce cyclonic flows and modify air-sea heat fluxes over the Atlantic basins through wind-evaporation-SST (WES) feedback[6,9,10]. These two pathways interact dynamically, with their relative strength modulated by the AMV-PDV phase relationship, as demonstrated by both observational analyses and idealized model pacemaker experiments: in-phase conditions favor Pacific-to-Atlantic influence, while out-of-phase conditions enhance Atlantic-to-Pacific control[11].

These Atlantic-Pacific interactions, while robust in observational records, present significant challenges for climate models. Mean state Sea Surface Temperature (SST) biases play a key role in misrepresenting decadal Atlantic-Pacific interactions by altering tropical convection thresholds and atmospheric bridges between ocean basins[2,12]. Studies of CMIP5 models showed that Atlantic and Pacific mean state SST biases independently weaken the CEP cooling response, with their combined effects explaining 89% of the reduced cooling compared to observations[2]. Despite improved representation of mean state and decadal variability in CMIP6 models, systematic biases persist in simulating Atlantic-Pacific teleconnections[4,13].

Precipitation variability plays a central role in modulating basin interactions through its effects on atmospheric circulation. Observational analyses show that equatorial Pacific precipitation anomalies modify inter-basin interactions, associated with alterations in tropical Walker circulation patterns[14]. Model representation of these processes varies widely, with Atlantic-to-Pacific response differences arising from biases in simulating moist static energy transport from the tropical Atlantic surface to the upper troposphere[4]. The phase relationship between PDV and AMV introduces additional complexity in precipitation responses. During in-phase conditions, suppressed Atlantic convection allows Pacific forcing to dominate, while enhanced convection during out-of-phase conditions enables Atlantic influence on Pacific variability[11].

Recent studies suggest that low cloud-SST feedback may affect basin interactions by modifying Pacific decadal SST variability. Cloud-locking experiments in CESM1 reveal that subtropical Northeastern Pacific low cloud alters local SST variability and subsequence influences basin-wide circulation through WES feedback[15]. The strength of low cloud-SST feedback in the subtropical Southeast Pacific correlates with Eastern Pacific multi-decadal SST variability across models through coupled cloud-radiative and circulation effects[16]. These results indicate that model biases in convective-scale physical processes may contribute to systematic errors in simulating Atlantic-Pacific teleconnections.

While previous studies have primarily focused on diagnosing individual pathway biases, particularly in the Atlantic-to-Pacific direction, a comprehensive evaluation of bidirectional Atlantic-Pacific interactions across climate models remains lacking[2,12,13,17]. Understanding how models represent these two-way interactions is crucial for improving climate predictions and reducing systematic biases. Additionally, given the importance of cloud processes in modulating basin interactions through their effects on SST variability and atmospheric circulation, investigating how different representations of cloud-scale dynamics influence these teleconnections represents an important research direction. The recent development of the Energy Exascale Earth System Model's multi-scale modeling framework (E3SM-MMF; also known as super-parameterization) alongside its traditional parameterized counterpart (E3SMv2) offers valuable insights into this question. E3SMv2 represents a state-of-the-art Earth system model with improved atmospheric dynamics and cloud microphysics[18], while E3SM-MMF replaces traditional sub-scale parameterizations with embedded cloud-resolving models, offering more realistic simulations of cloud processes and their interactions with large-scale dynamics[19]. Recent advances in GPU-acceleration have enabled a 65-year simulation of the fully coupled E3SM-MMF, allowing unprecedented investigation of decadal-scale tropical variability in a super-parameterized framework.

This study examines patterns of Atlantic-Pacific interactions across 27 CMIP6 models, investigating how these models represent bidirectional teleconnections compared to observational evidences. By analyzing E3SMv2 and E3SM-MMF alongside the CMIP6 ensemble, we further explore how different treatments of cloud-scale processes might influence these basin interactions. Our analysis provides new insights into systematic biases in model representations of Atlantic-Pacific interactions and evaluates whether the coupled super-paramterized model framework could offer pathways for improvement.

# Results

## Tropical Pacific-driven Dominance in CMIP6 Models

To charactrize the relationship of Atlantic-Pacific interactions, we conduct the linear regression analysis of low-pass filtered (10-year) global SSTA onto the NTA index (Fig. 1). The figure exhibits potential biases in CMIP6 models' representation of Atlantic-Pacific interactions, a challenge that has been widely documented in previous studies[2,4,12,13,17]. Both ERSST and HadISST observations reveal that NTA warming generates an inter-hemispheric Atlantic dipole pattern coupled with Pacific cooling, resembling La Niña-like conditions (Fig. 1a, 1b), consistent with findings from multiple observational analyses and targeted GCM pacemaker experiements [5,6,20,21]. Instead of the observed equatorially asymmetric Atlantic structure (Northern Hemisphere warming and Southern Hemisphere cooling) and Eastern Pacific cooling, CMIP6-MMM exhibits basin-wide warming across both oceans, with an El Niño-like pattern in the Pacific (Fig. 1c). This systematic model bias in representing Atlantic-Pacific teleconnections has been identified across multiple generations of climate models[2,13,22]. This characteristic appears consistently in the model ensemble - inter-model consistency tests show that almost all 27 CMIP6 models simulate this warming response (Fig. S1).  While the magnitude varies across models, the spatial structure of this warming pattern remains consistent, indicating a systematic feature across CMIP6 models on decadal Atlantic-Pacific teleconnections.

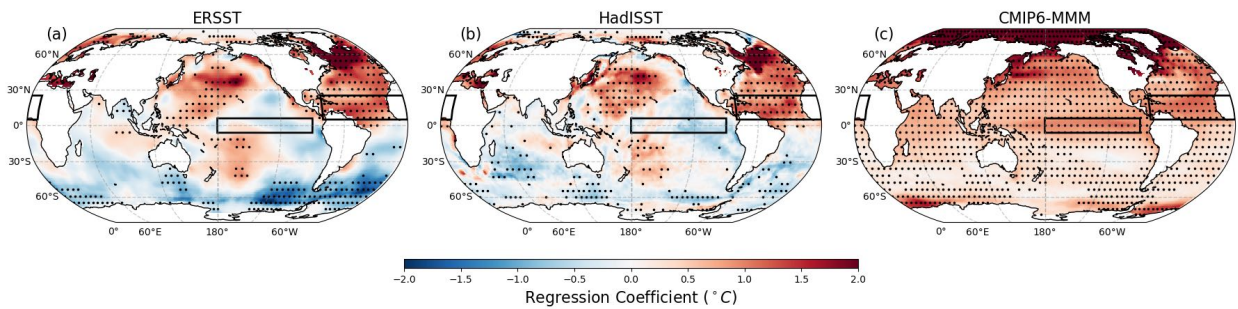

**Figure 1: Regression patterns of sea surface temperature anomalies onto the North Tropical Atlantic index.** SSTA regression coefficients for (a) ERSST, (b) HadISST, and (c) CMIP6 multi-model mean during 1950-2015. Stippling indicates statistical significance at 90% confidence level based on Student's t-test with effective degrees of freedom for observations, and 90% model agreement on sign for CMIP6-MMM.

**Contrasting Walker Circulation Responses in Observations versus Models.** Analysis of tropical and extratropical pathways reveals fundamental differences between observed and modeled Atlantic-Pacific interactions, particularly in their basin dominance patterns. Investigation of tropical pathways using linear regressions of equatorial (5°S-5°N) vertical velocity onto both NTA and CTI shows reversed Walker circulation patterns in ERA5 reanalysis (Figs. 2a and 2c). Following the mechanistic framework established by ref.[5], we interpret the observational patterns as a sequence of coupled ocean-atmosphere interactions: During NTA warming, ERA5 exhibits ascending motion over the Atlantic that triggers secondary ascending motion over the Indo-Western Pacific. These ascending motions induce compensating subsidence over the CEP (Fig. 2a). The circulation patterns may intensify Pacific easterlies and enhances surface evaporative cooling through WES feedback. During CTI warming, the tropical circulation reverses - ascending motion over the Pacific induces subsidence over both Atlantic and Western Pacific sectors (Fig. 2c). These out-of-phase circulation patterns indicate separate processes governing each direction of interaction.

CMIP6 models, however, fail to capture this observed asymmetry in tropical circulation responses (Figs. 2b and 2d). While models accurately reproduce the Pacific-to-Atlantic pathway, where CTI warming generates ascending motion driving Atlantic subsidence (Fig. 2b and Fig. S2), they misrepresent the Atlantic-to-Pacific connection. In contrast to observations, their response to Atlantic warming shows ascending motion over CEP (Fig. 2d and Fig. S3), displaying a circulation pattern that matches the Pacific-to-Atlantic response but with opposite sign to observations. This unrelatistic simulated Walker circulation response potentially weakens Pacific trade winds and reduces surface evaporative cooling, consistent with an incorrect East Pacific warming pattern. The unrealistically consistent circulation patterns across both Pacific and Atlantic forcing scenarios indicate models overestimate Pacific influence in decadal Atlantic-Pacific interactions.

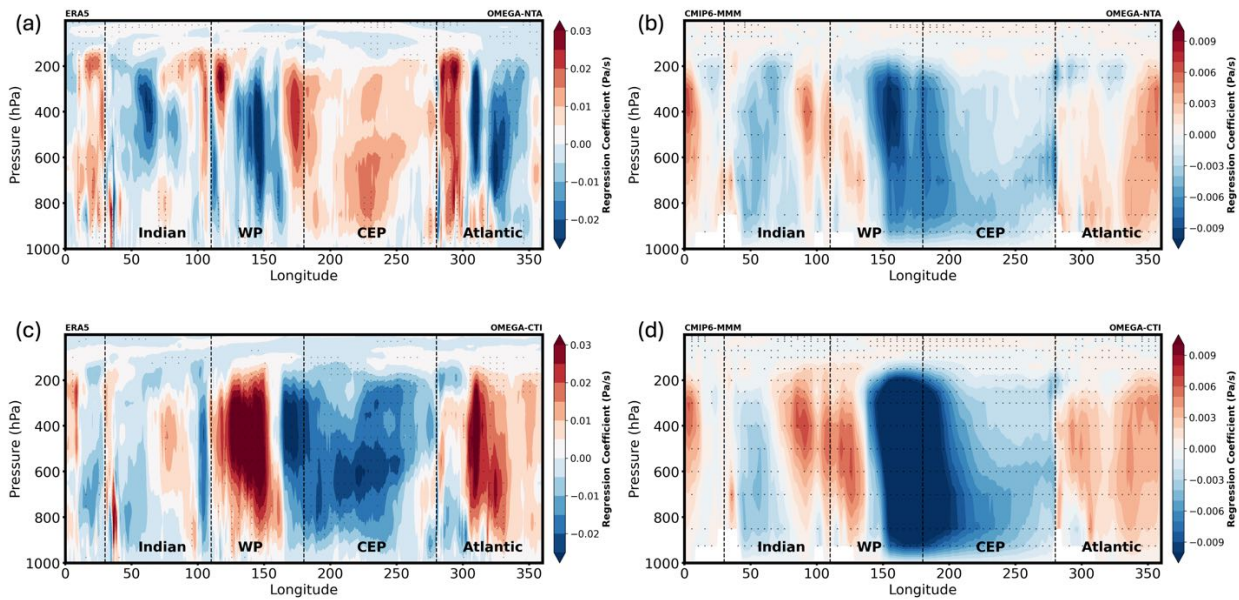

**Figure 2: Walker circulation responses to tropical Atlantic and Pacific forcing.** Regression coefficients of equatorial (5°S-5°N) vertical pressure velocity (OMEGA) onto the NTA index (a,b) and Cold Tongue Index (c,d) for ERA5 (left; ±0.03 Pa/s) and CMIP6-MMM (right; ±0.009 Pa/s). Positive values (red) indicate descending motion; negative values (blue) indicate ascending motion. Statistical significance as in Figure 1.

**Divergent Extratropical Teleconnection Patterns in Models and Observations.** Linear regression of Z200 against basin SSTs further supports Pacific dominance bias across CMIP6 models from the extratropical teleconnection (Fig. 3). During NTA warming, ERA5 shows upper-level convergence centered over the CEP (180°E-240°E), indicating an enhanced Walker circulation (Fig. 3a). The upper-level convergence triggers poleward-propagating Rossby waves, generating an anticyclonic anomaly over the Aleutian region (~60°N) followed by a cyclonic circulation along western North America (~60°N), resembling a negative PNA-like pattern and a symmetric wave train in the Southern Hemisphere. In response to CTI warming, the atmosphere displays a Gill-type response with equatorial symmetric cyclonic pairs (Fig. 3c). The tropical warming triggers a PNA wave train from the tropical Pacific to the North Atlantic, weakening the North Atlantic subtropical high (20°N-40°N) and potentially northeastern trade winds, which may reduce latent heat flux in the tropical North Atlantic. Similarly, the PSA pattern in the Southern Hemisphere weakens the South Atlantic subtropical high and its surface winds. These hemispherically symmetric atmospheric bridges enable Pacific warming to generate basin-wide Atlantic warming[6].

The CMIP6-MMM captures the atmospheric response to CEP warming (Fig. 3d), reproducing both the Gill-type response and subsequent extratropical wave trains. However, for NTA forcing (Fig. 3b), it fails to reproduce the distinct centers of positive and negative Z200 anomalies seen in ERA5, instead generating a spatially uniform pattern of positive anomalies across the domain (Fig. 3b). This asymmetric skill is associated with SSTA and Walker circulation anomaly, reflecting models' deficiency in simulating Atlantic-to-Pacific teleconnections. While inter-model differences weaken the CMIP6-MMM signal, examination of individual models confirms this bias as a common feature across the model ensemble (Fig. S4 and Fig. S5).

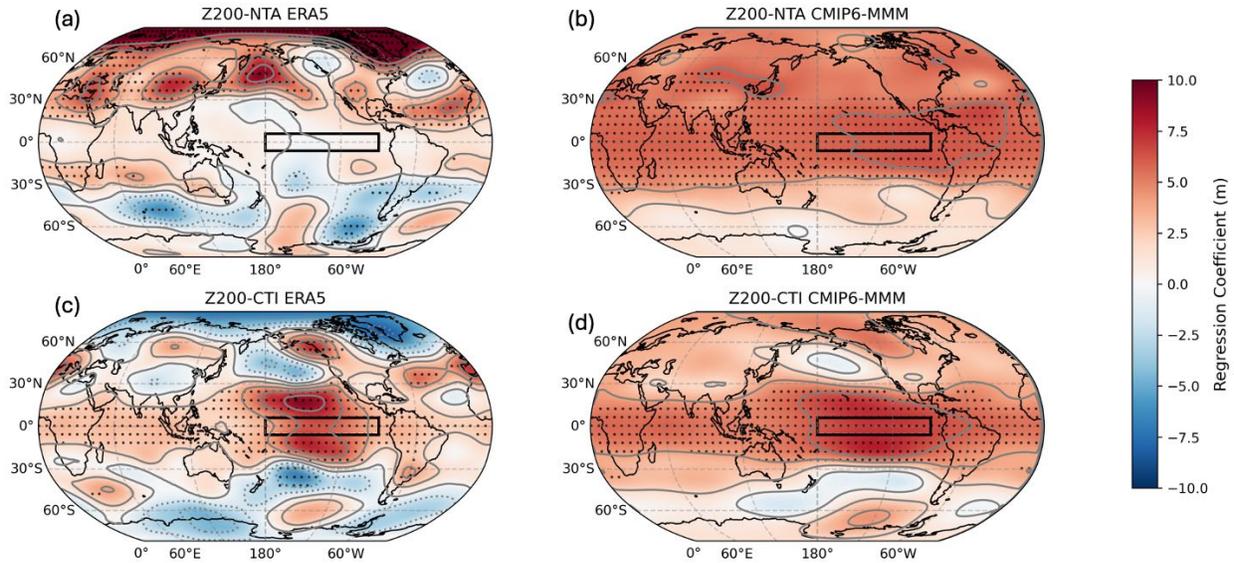

**Figure 3: Extratropical teleconnection patterns in response to tropical Atlantic and Pacific forcing.** Regression patterns of 200-hPa geopotential height (Z200) onto the NTA index (a,b) and CTI (c,d) for ERA5 (left) and CMIP6-MMM (right). Contour interval is 2m. Red (blue) shading indicates positive (negative) height anomalies. Statistical significance as in Figure 1.

**Pattern correlation analysis as a useful summary for systematic model bias.** To evaluate individual model performance in representing basin interactions, we calculate pattern correlations between Atlantic-forced and Pacific-forced regression patterns. The analysis focuses on two domains: the vertical velocity field (180°E-360°E, 5°S-5°N) for tropical pathways, and the Z200 responses (180°E-360°E, 60°S-60°N) for extratropical Rossby wave propagation paths.

Pattern correlation analysis of Walker circulation reveals contrasting basin relationships between observations and models (Fig. 4). ERA5 shows a negative correlation (-0.76) between Atlantic-forced and Pacific-forced circulation patterns, consistent with the observed distinct atmospheric responses to each basin's forcing. The CMIP6 multi-model mean shows an opposite response with a positive correlation (0.50 ± 0.38), with 56% of models showing correlations above 0.50 and only 3 of 27 models reproducing the observed negative correlation (Fig. 4a). This systematic bias extends to extratropical domains - while ERA5 exhibits negative pattern correlation in Z200 anomalies between Atlantic-forced and Pacific-forced conditions, 93% of CMIP6 models show strong in-phase relationships ($r > 0.6$) (Fig. 4b).

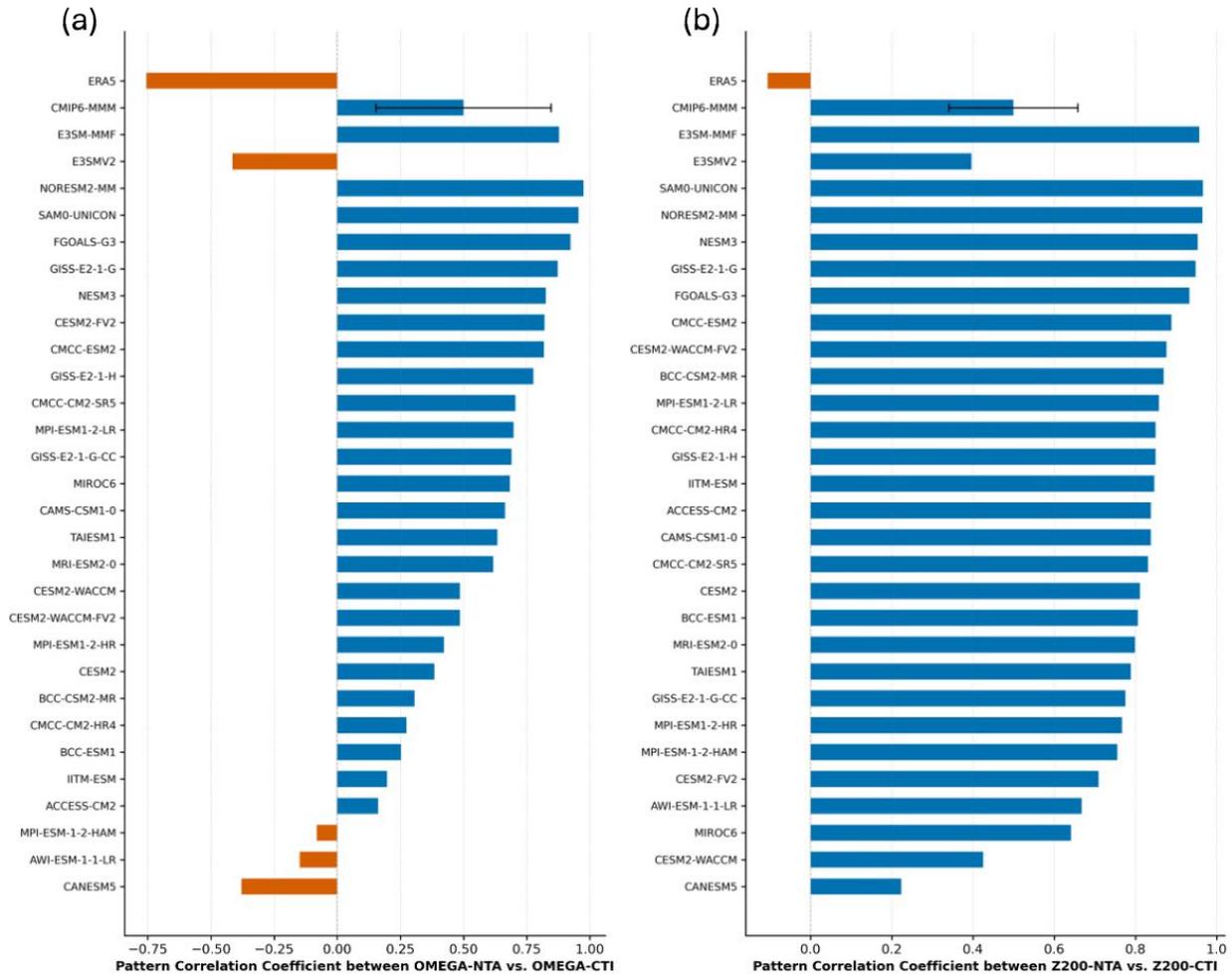

**Figure 4: Contrasting basin dominance between observations and models.** Pattern correlations between NTA-forced and CTI-forced responses for (a) equatorial vertical velocity (180°E-360°E, 5°S-5°N) and (b) Z200 (180°E-360°E, 60°S-60°N). Bars show correlation coefficients for ERA5 and models. For CMIP6, the bar shows multi-model mean with error bars indicating one standard deviation across 27 models.

In summary, multiple lines of evidence from tropical and extratropical pathways reveal opposite basin dominance between observations and CMIP6 models. While observations show tropical Atlantic control of Pacific variability on decadal timescales, CMIP6 models display tropical Pacific dominance in basin interactions. Previous studies have documented models' deficiency in simulating Pacific responses to Atlantic forcing through both SST patterns and atmospheric circulation changes[12,17]. Our examination of bidirectional teleconnections demonstrates that models systematically overemphasize Pacific-to-Atlantic pathways, undermining their ability to capture key Atlantic forcing processes, including Walker circulation changes driven by Atlantic multidecadal oscillations[20] and atmospheric bridges linking tropical Atlantic SST anomalies to

Pacific conditions. These findings highlight the need to improve model physics to better represent the observed capability of Atlantic variability to drive Pacific changes.

## A Total Performance Metric for the Realism of CMIP6 Models' Atlantic-Pacific Interaction

Based on the Total Performance Metric (TPM) described in the Methods section, we classified CMIP6 models into two categories. Models with TPM<1.0 were classified as high-skill (4 models: CanESM5, AWI-ESM-1-1-LR, CESM2-WACCM, and MPI-ESM-1-2-HAM), reproducing observational-like teleconnection patterns. Models with TPM>1.5 were classified as low-skill (15 models), indicating unrealistic Pacific-driven teleconnections. This classification reveals that most CMIP6 models exhibit significant biases in representing Atlantic-Pacific interactions (see Supplement Table 1 for model details).

Linear regression of global SSTA onto the NTA index confirms the usefulness of TPM for contrasting Pacific responses between high-skill and low-skill model groups (Fig. 5). The high-skill MMM shows warming signal in the WP, though it produces equatorial CEP warming (CTI region value of 0.58°C) that opposes the observed cooling pattern (Fig. 5a). The low-skill MMM exhibits a substantially stronger El Niño-like Pacific response (CTI region value of 1.08°C), characterized by equatorial warming extending from eastern to western Pacific (Fig. 5b). Unlike the observed meridional dipole pattern in the Atlantic, both high-skill and low-skill models simulate uniform warming across the Atlantic basin in response to NTA warming.

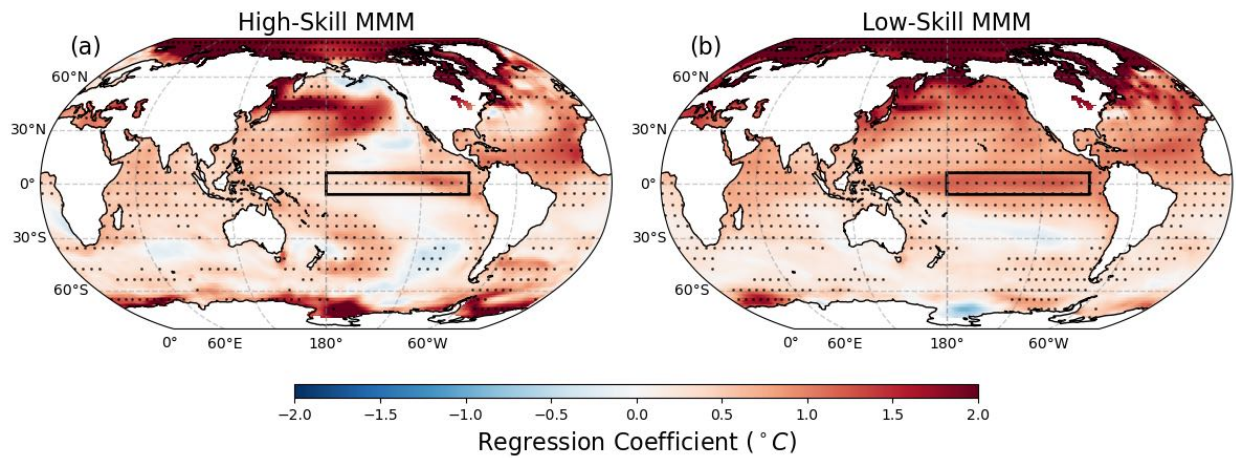

**Figure 5: SST regression patterns in high-skill versus low-skill models.** Same regression analysis as Figure 1 but for (a) high-skill multi-model mean (TPM < 1.0) and (b) low-skill multi-model mean (TPM > 1.5). Statistical significance as in Figure 1.

Walker circulation anomaly regressions reveal systematic spatial patterns in model responses (Fig. 6). During NTA warming, the high-skill MMM maintains the fundamental circulation structure while exhibiting two key differences from observations: a 20° westward displacement of the ascending branch and reduced vertical velocity (approximately 30% of observed magnitude), consistent with findings from previous studies[2,4,5]. This spatial shift in circulation appears associated with anomalous ascending motion in the central equatorial Pacific, which could contribute to the equatorial Pacific warming bias observed in the models (Fig. 6c). Under CTI warming conditions, the high-skill MMM exhibits a westward displacement of the ascending branch toward the western Pacific (150-200°E), with weakened but correctly-signed Atlantic subsidence (Fig. 6d). In contrast, the low-skill MMM shows limited sensitivity to the forcing conditions, producing nearly identical circulation patterns in both cases: enhanced western Pacific ascending motion coupled with Atlantic subsidence, suggesting a systematic bias toward Pacific-dominated variability (Fig. 6e and 6f).

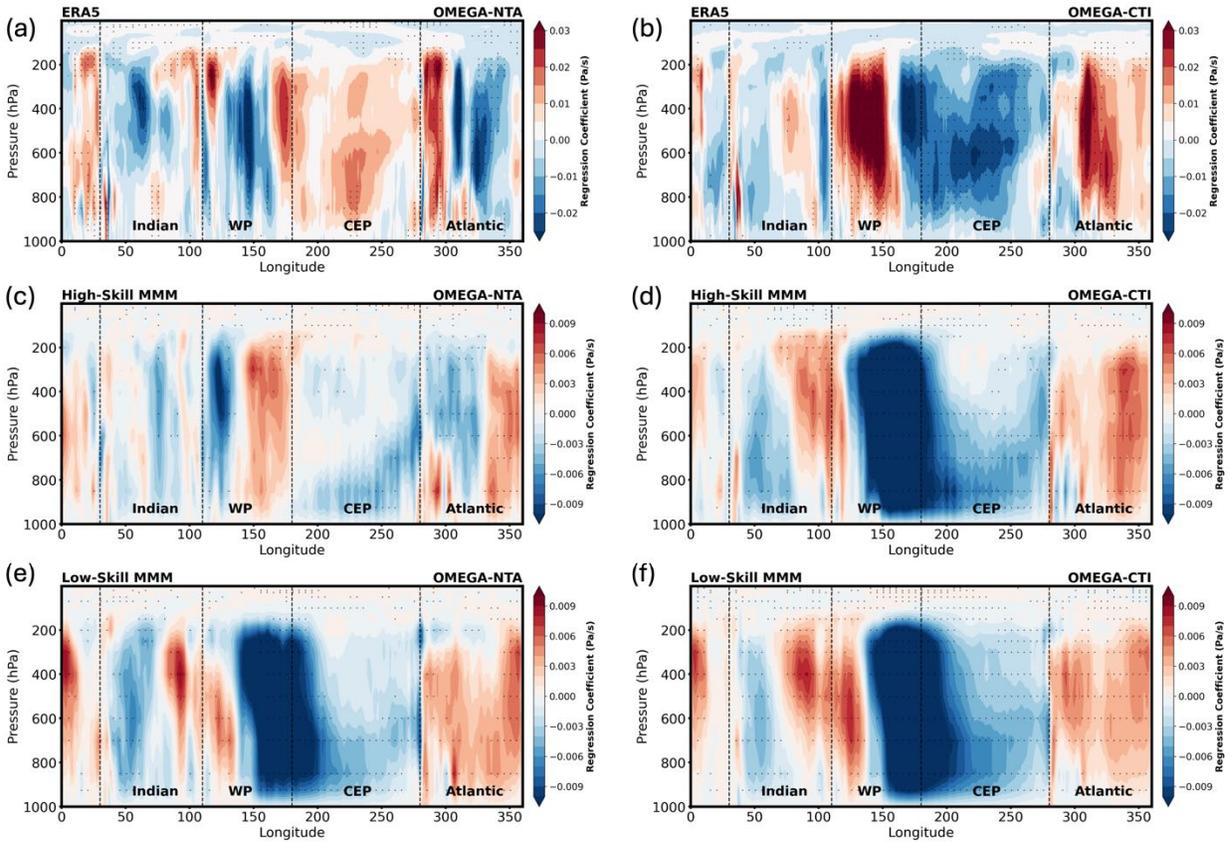

**Figure 6: Walker circulation responses across model skill groups.** Regression coefficients of equatorial vertical pressure velocity onto NTA index (left) and CTI (right) for (a,b) ERA5 (±0.03 Pa/s), (c,d) high-skill multi-model mean (TPM < 1.0; ±0.009 Pa/s), and (e,f) low-skill multi-model mean (TPM > 1.5; ±0.009 Pa/s). Statistical significance as in Figure 1.

Equatorial precipitation biases (5°S-5°N) are unsurprisingly associated with the Walker circulation biases identified above (Fig. 7a and 7b). During NTA warming, ERA5 exhibits precipitation maxima at 330°E (Western Atlantic) and 150°E (WP), separated by suppressed precipitation (300°E-180°E). The high-skill MMM maintains positive Atlantic precipitation despite the 20° westward displacement, preserving the observed teleconnection direction. The low-skill MMM generates negative Atlantic precipitation coupled with enhanced CEP precipitation, indicating reversed tropical teleconnection pathways (Fig. 7a). Under CTI warming, both groups display westward-shifted WP precipitation maxima relative to ERA5, maintaining a weaker but correctly signed CEP and subsequent Atlantic responses (Fig. 7b).

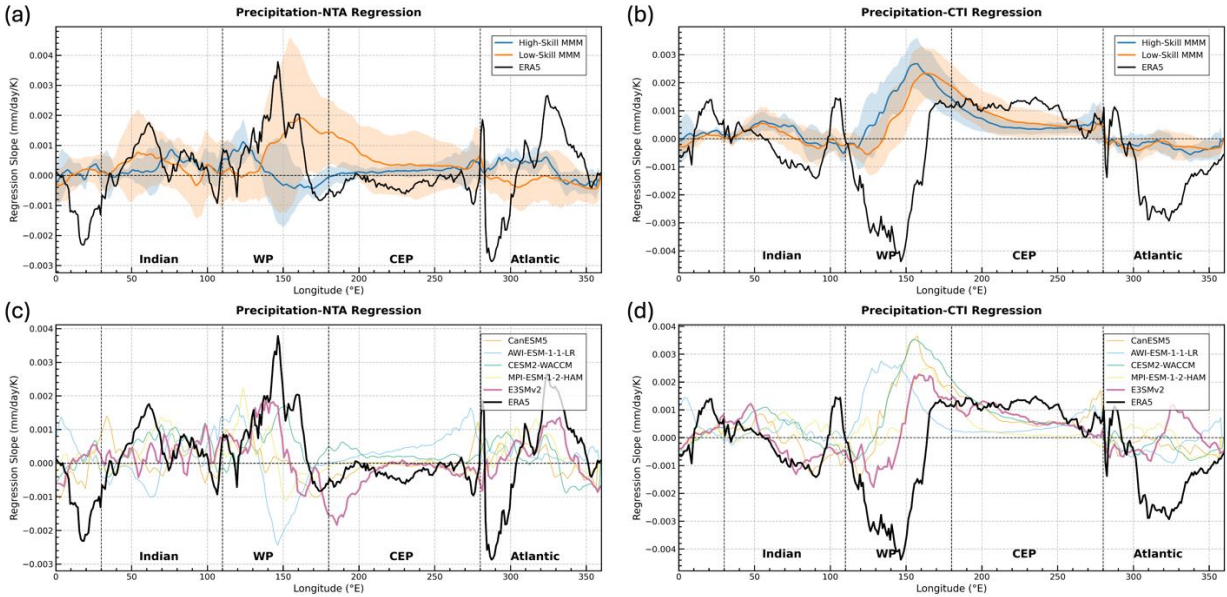

**Figure 7: Equatorial precipitation responses to basin forcing.** Regression of equatorial precipitation anomalies (mm/day/k) averaged between 5°S-5°N onto (a,c) NTA and (b,d) CTI indexes during 1950-2014. Upper panels show ERA5 (black), high-skill multi-model mean (blue with ±1 standard deviation shading), and low-skill multi-model mean (orange with ±1 standard deviation shading). Lower panels show ERA5 (black) and individual high-skill models.

The Z200 regressions reveal distinct wave train patterns between model groups (Fig. 8). During NTA warming, the high-skill MMM shows a southwestward displacement of the positive geopotential height center from Eastern Atlantic (25°N-35°N, 320°E-350°E) to tropical Western Atlantic (5°N-15°N, 300°E-330°E), which is consistent with tropical pathway analysis. This shifted high-pressure system induces a reduced positive geopotential height anomaly in the CEP (5°N-15°N , 180°E-200°E ), generating a negative PNA-like wave train response across North America. Under CTI warming, the group produces symmetric wave trains in both hemispheres, with PNA and PSA patterns connecting Pacific anomalies to the Atlantic basin. The low-skill MMM exhibits excessive positive Z200 anomalies in the tropical Pacific region, indicating unrealistic Pacific dominance over Atlantic variability.

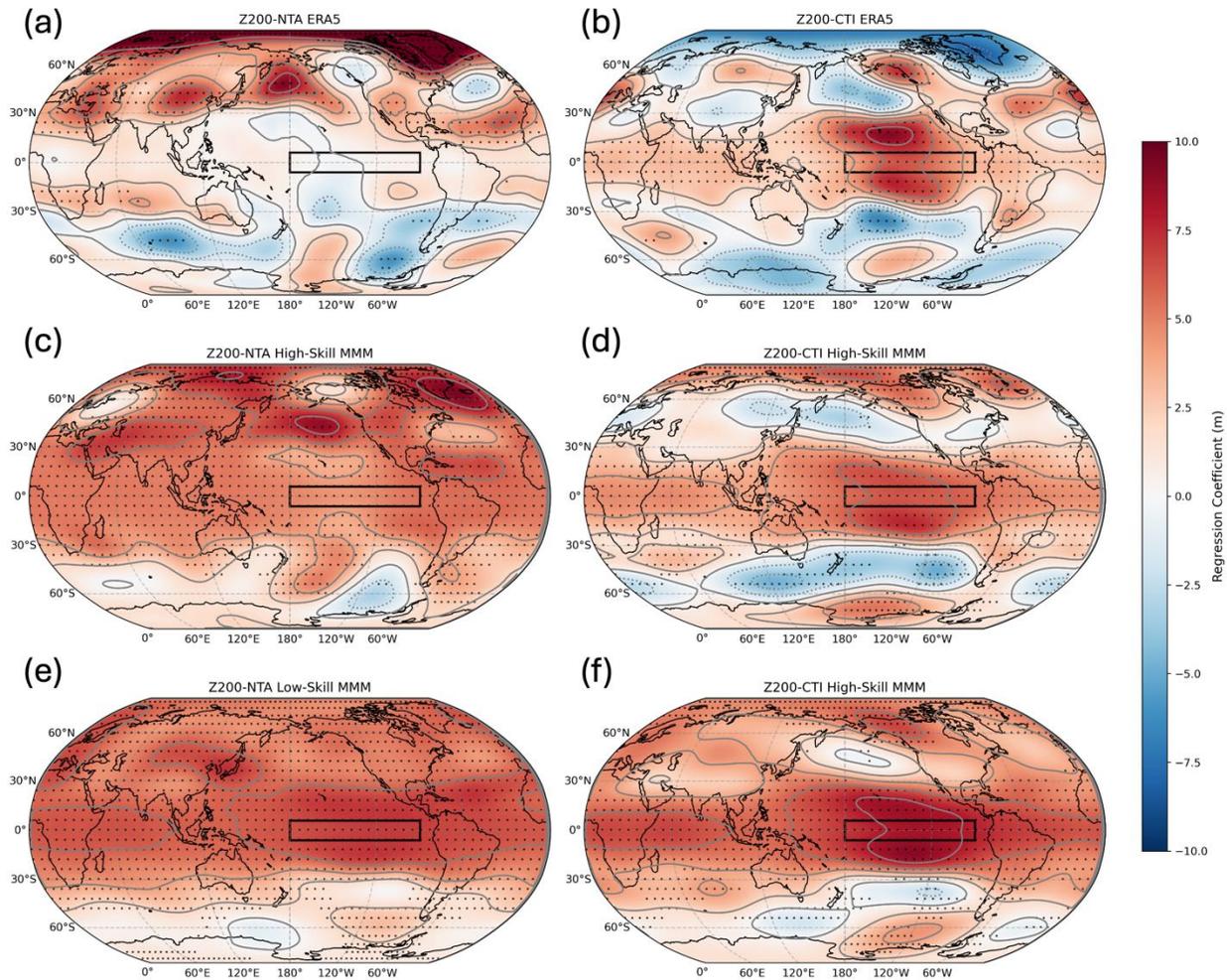

**Figure 8: Extratropical responses across model skill groups.** Regression patterns of Z200 onto NTA index (left) and CTI (right) for (a,b) ERA5, (c,d) high-skill multi-model mean (TPM < 1.0), and (e,f) low-skill multi-model mean (TPM > 1.5). Contour interval is 2m. Red (blue) shading indicates positive (negative) height anomalies. Statistical significance as in Figure 1.

Our analyses characterize systematic differences in how models represent tropical basin interactions. High-skill models maintain the observed teleconnection directionality despite showing consistent spatial displacements in their local responses. In contrast, low-skill models demonstrate reversed inter-basin interactions under Atlantic forcing and excessive Pacific influence. These distinct model behaviors motivate further examination of high-skill models to better understand the features that enable more realistic tropical basin interactions.

## Characteristics of high-skill models on Atlantic-Pacific interactions

High-skill models (CanESM5, AWI-ESM-1-1-LR, CESM2-WACCM and MPI-ESM-1-2-HAM) reproduce the observed tripolar SST pattern in the North Atlantic and warming pattern in the WP, reflecting their improved representation of NTA warming response (Fig. 9). However, their representation of SSTA in the CEP shows model-dependent variations. Specifically, CanESM5 and AWI-ESM-1-1-LR simulate an observationally consistent off-equatorial cooling pattern in the CEP, characterized by equatorially symmetric cooling bands in both hemispheres. CESM2-WACCM and MPI-ESM-1-2-HAM, in contrast, display asymmetrical responses, with prominent warming in the Southern Hemisphere and cooling in the Northern Hemisphere. Besides, models are varied for the equatorial warming response in the CEP. The AWI-ESM-1-1-LR shows a strong warming bias (>2°C) in the Pacific cold tongue region, with other three models show weakened warming responses (~0.5°C).

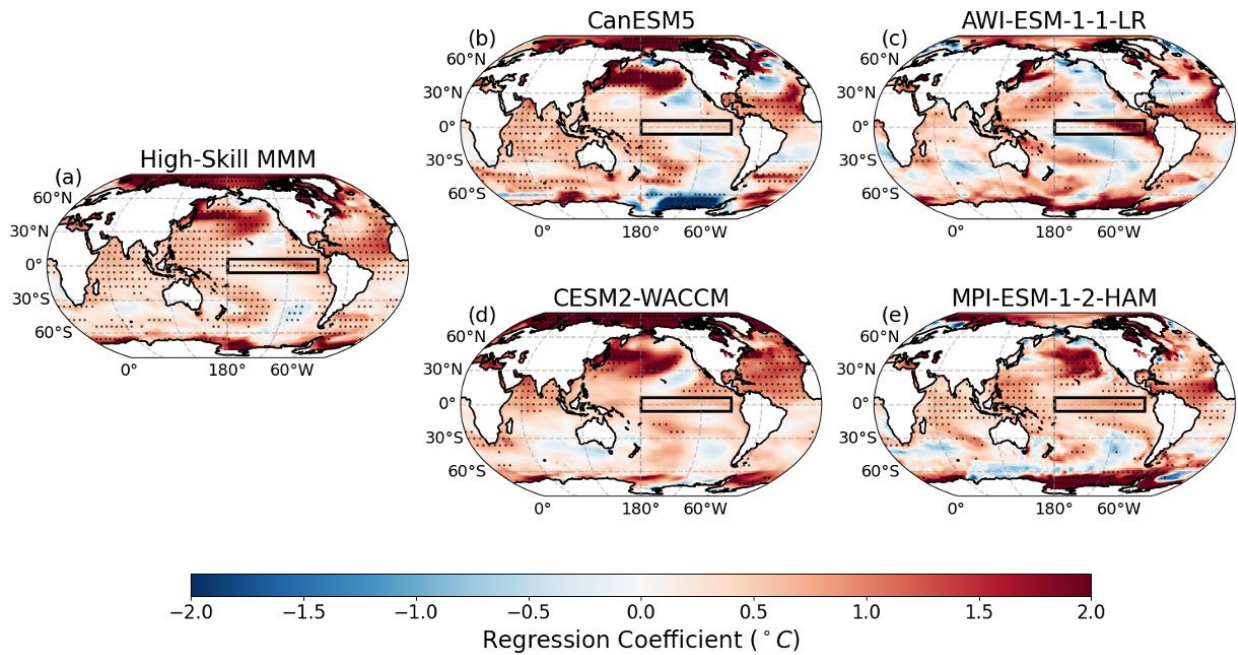

**Figure 9: SST responses to Atlantic forcing in high-skill models.** Regression coefficients of SSTA onto the NTA index for (a) high-skill multi-model mean, (b) CanESM5, (c) AWI-ESM-1-1-LR, (d) CESM2-WACCM, and (e) MPI-ESM-1-2-HAM. Statistical significance as in Figure 1.

High-skill models exhibit varying Walker circulation responses that correspond to their distinct SSTA patterns (Fig. 10). During NTA warming, they show a consistent but biased local response, with ascending motion shifted 20° westward and reduced to 1/3 of ERA5 magnitude, matching the high-skill MMM characteristics described above. The models diverge in their representation of Pacific vertical motion and precipitation, producing varied SSTA regression patterns in the CEP.

For example, AWI-ESM-1-1-LR exhibits an unrealistic strong ascending response and positive precipitation anomaly in the Pacific cold tongue region during NTA warming (Fig. 10e). The model shows westward-shifted Walker circulation anomalies, characterized by strong ascending motion in the eastern equatorial Pacific (250°E -280°E) and descending motion in the western equatorial Pacific (150°E -180°E). These circulation biases may be connected through a sequence of atmospheric and oceanic processes: Atlantic warming appears to enhance Indo-Western Pacific convection (100°E -130°E) as described in[5], which could strengthen subsidence in the western equatorial Pacific. This altered circulation pattern may contribute to weakened easterly winds in the eastern equatorial Pacific, potentially enabling warming through wind-evaporation-SST feedback mechanisms. Studies have shown that CMIP6 models exhibit weak oceanic wave dynamics[23,24], which could limit their ability to capture cooling effects in the eastern equatorial Pacific.

Among high-skill models, CanESM5 best reproduces the observed Walker circulation and zonal precipitation patterns, capturing both the Atlantic-forced subsidence in the CEP and precipitation maxima in the WP (Fig. 10c). However, during CTI warming, all high-skill models misrepresent the Pacific response by displacing the convective center westward from CEP to WP. This systematic Walker cell shift prevents proper simulation of the observed negative precipitation anomalies in the CEP. MPI-ESM-1-2-HAM exhibits the most pronounced version of this bias, concentrating updrafts in the WP. This concentration weakens both CEP and Atlantic subsidence, producing an unrealistic Pacific-to-Atlantic teleconnection pattern.

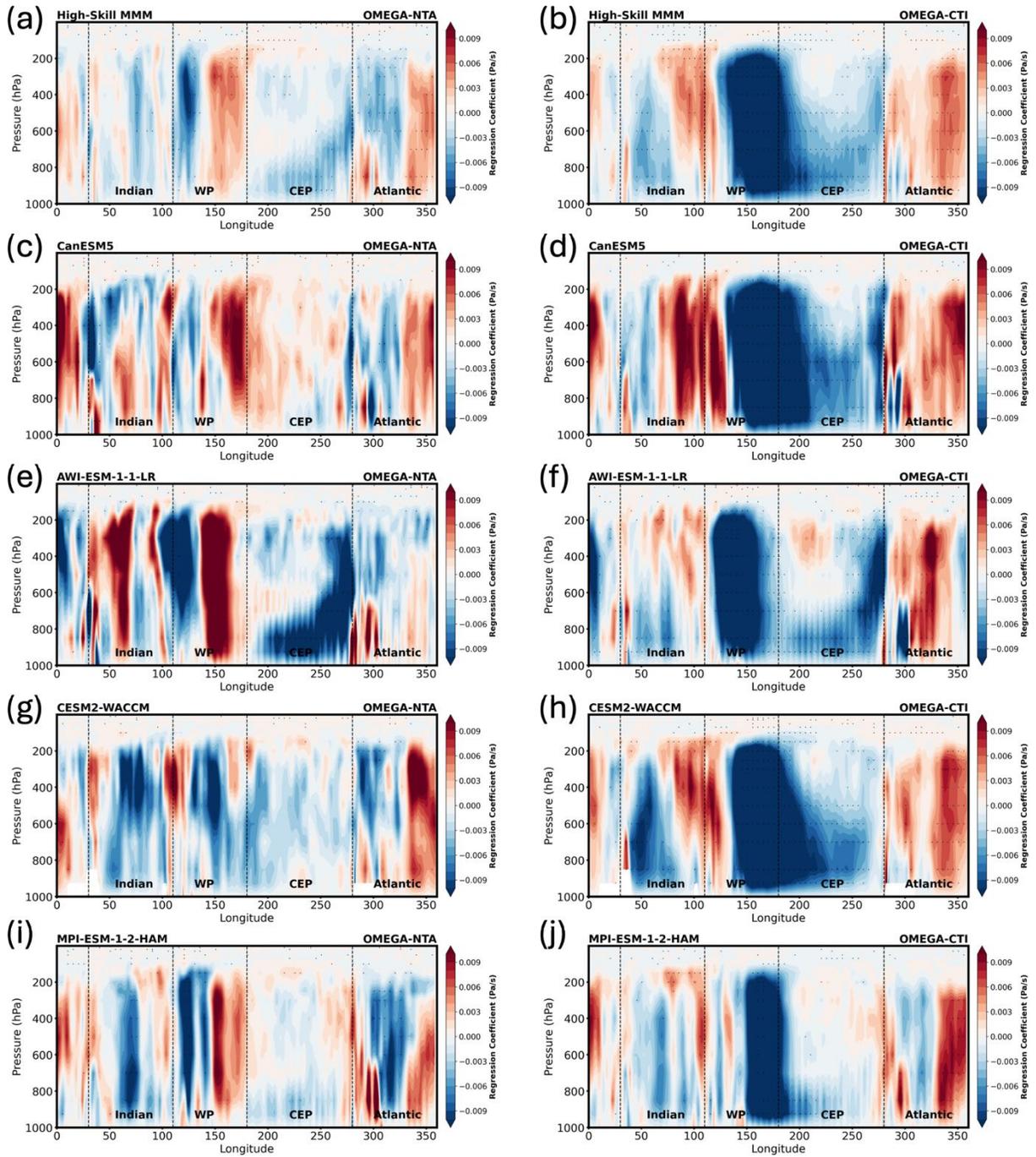

**Figure 10: Walker circulation responses in individual high-skill models.** Regression coefficients of equatorial vertical pressure velocity onto NTA index (left) and CTI (right) for (a,b) high-skill multi-model mean and individual high-skill models: (c,d) CanESM5, (e,f) AWI-ESM-1-1-LR, (g,h) CESM2-WACCM, and (i,j) MPI-ESM-1-2-HAM. Statistical significance as in Figure 1.

High-skill models simulate upper-tropospheric circulation responses with systematic biases in upper tropospheric geopotential height anomalies. Under both NTA and CTI warming, these models produce stronger positive height anomalies across the tropical region compared to ERA5, with this bias particularly evident during NTA forcing (Fig. 11). This enhanced tropical response generates a diffuse positive pattern in their multi-model mean, partially masking the wave train signals in both basins.

Despite this common tropical bias, the models reproduce key wave train characteristics. During NTA warming, they simulate a negative-PNA wave train pattern extending from the tropical Pacific to the Atlantic in the Northern Hemisphere. This wave train structure aligns with their simulated North Atlantic tripolar SSTA pattern, featuring alternating centers of high and low pressure. The strongest signals appear as a negative anomaly over the North Pacific and a positive anomaly over North America, consistent with the models' representation of basin-scale SST responses.

The wave train patterns vary across models in position and intensity, corresponding to their distinct CEP SSTA responses. CanESM5, CESM2-WACCM, and AWI-ESM-1-1-LR capture the negative-PNA wave train response, with wave train positions reflecting their CEP cooling patterns. MPI-ESM-1-2-HAM generates a westward-shifted wave train structure, corresponding to its biased western Pacific SSTA response (Fig. 11i). During CTI warming, the models simulate the Gill-type response with symmetrical PNA and PSA wave train patterns in both hemispheres, though the tropical height bias modifies their relative amplitudes.

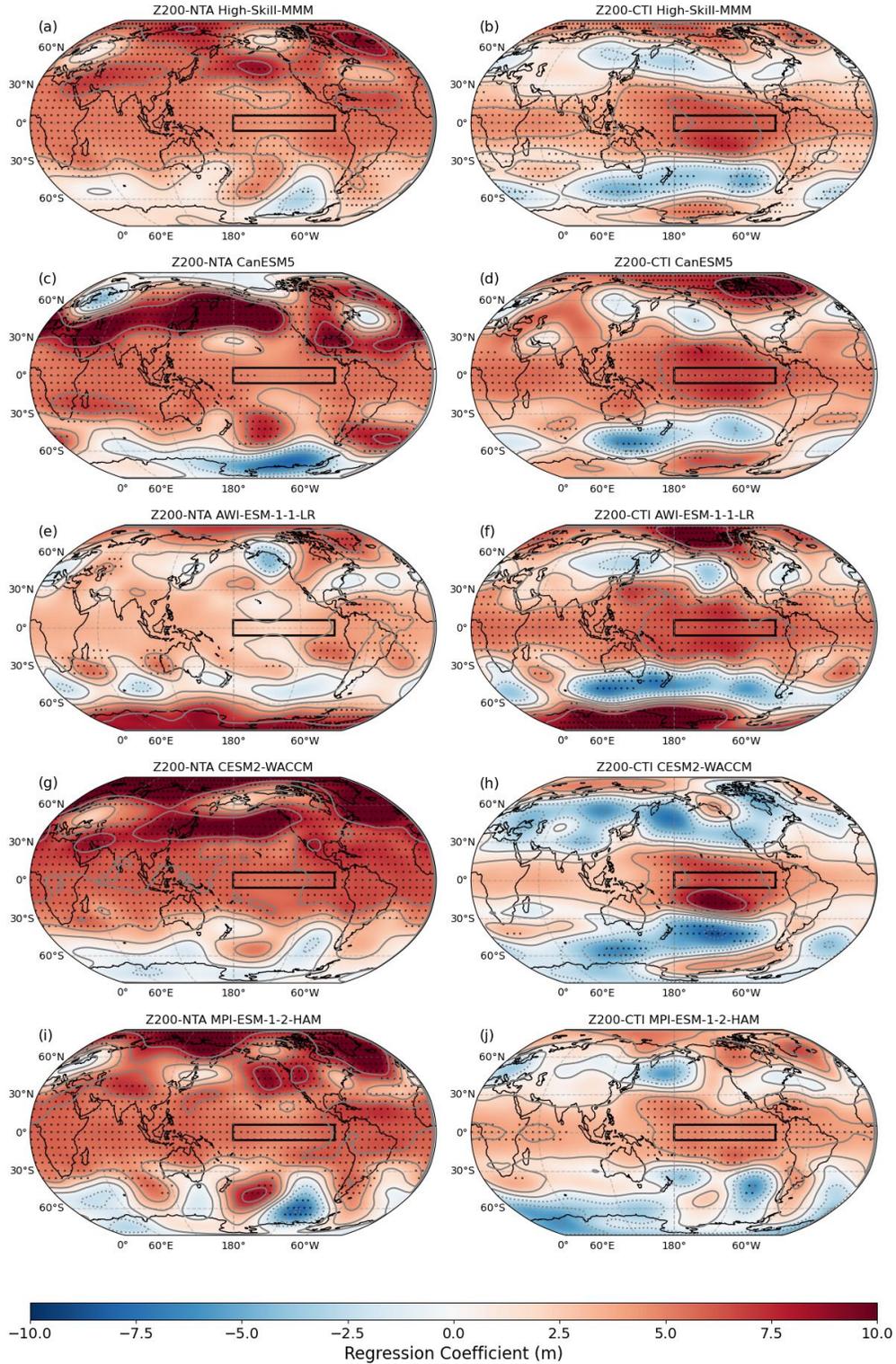

**Figure 11: Extratropical teleconnection patterns in high-skill models.** Regression patterns of Z200 onto NTA index (left) and CTI (right) for (a,b) high-skill multi-model mean and individual models: (c,d) CanESM5, (e,f) AWI-ESM-1-1-LR, (g,h) CESM2-WACCM, and (i,j) MPI-ESM-1-2-HAM. Statistical significance as in Figure 1.

# Discussion

## Atlantic versus Pacific Dominance: Contrasting Observations and Models

Our diagnostic analysis of 27 CMIP6 models reveals systematic biases in representing decadal Atlantic-Pacific teleconnections, with important implications for climate prediction. Most CMIP6 models reverse the observed direction of Atlantic-to-Pacific interactions on decadal timescales, with Pacific variability dominating Atlantic responses rather than the observed Atlantic influence on Pacific climate. By developing a TPM metric that rewards diversity between tropical Walker circulation and extratropical wave responses to Atlantic versus Pacific control, we identified two distinct model groups.

High-skill models (TPM < 1.0) capture the general structure of Atlantic-Pacific teleconnections with a secondary symptom of systematic westward shifts in convective centers within basins, resulting in weaker La Niña-like Pacific responses to Atlantic warming. In contrast, low-skill models (TPM > 1.5) exhibit Pacific-dominated responses regardless of the forcing basin, producing El Niño-like warming in association with Atlantic warming, opposite to observations.

## Contrasting Atlantic-Pacific Interactions in E3SM Configurations

While our primary analysis focused on CMIP6 models, we conducted additional comparative analysis between E3SMv2 and its multi-scale modeling framework (E3SM-MMF) to explore how different treatments of cloud-scale processes influence these basin interactions. This complementary examination provides valuable insights into the potential role of model physics in teleconnection representation.

E3SMv2 performs remarkably well in reproducing observed SSTA responses to NTA warming, capturing both the Atlantic meridional dipole and Pacific off-equatorial cooling patterns (Fig. 12c)—features that even the four high-skill CMIP6 models failed to simulate consistently. Its tropical circulation response during NTA warming correctly simulates ascending motion centers with proper spatial structure compared to other high-skill models, with a reduced bias magnitude of approximately 54% in vertical velocity relative to ERA5. Under CTI warming, E3SMv2 maintains the observed Walker circulation structure across basins, though with a 20° westward displacement of ascending motion and enhanced ascending anomalies in the tropical central Atlantic (320°E -330°E). In the extratropical pathway, E3SMv2 outperforms other high-skill models in reproducing the observed negative-PNA wave train and North Atlantic geopotential height anomalies during both NTA warming phase, though with amplified magnitude.

E3SM-MMF displays systematic biases characteristic of low-skill models in both tropical and extratropical pathways (Fig. 12d). The model produces Pacific-dominated responses regardless of the forcing region, with persistent ascending motion biases in the tropical Pacific and reversed Walker circulation responses to Atlantic warming. Its extratropical response shows Pacific-originating wave trains during both NTA and CTI warming, with Z200 anomalies twice the observed magnitude in the tropical band (30°S-30°N).

The contrasting performance between E3SMv2 and E3SM-MMF in representing decadal Atlantic-Pacific interactions raises important questions about the role of cloud-scale processes in these emergent large-scale dynamics. These results could be viewed as counterintuitive given previous findings at interannual timescales. For instance, ref.[25] demonstrated that coupled superparameterization improved the simulation of South Asian monsoon variability compared to traditional parameterization, suggesting benefits of explicit cloud-scale process treatments for large-scale dynamics. Their study also revealed that atmosphere-ocean coupling enhanced the interaction between small-scale and large-scale dynamics in the superparameterized framework. The substantially degraded performance of E3SM-MMF at decadal timescales may reflect the limited model retuning in its development due to computational limitations. Beyond this technical consideration, the representation of cloud processes influences tropical circulation patterns through multiple processes as documented by previous studies: cloud-radiation feedbacks affect tropical circulation through convective self-aggregation processes[28], shallow convection critically influences Walker circulation strength and tropical climate[29], and cloud thermodynamic processes substantially impact tropical-extratropical connections through their effects on deep convection and anvil cloud formation[30,31]. These findings suggest that the current configuration of E3SM-MMF may not optimally capture the complex interactions between cloud-scale processes and large-scale dynamics at decadal timescales, particularly in the context of Atlantic-Pacific interactions.

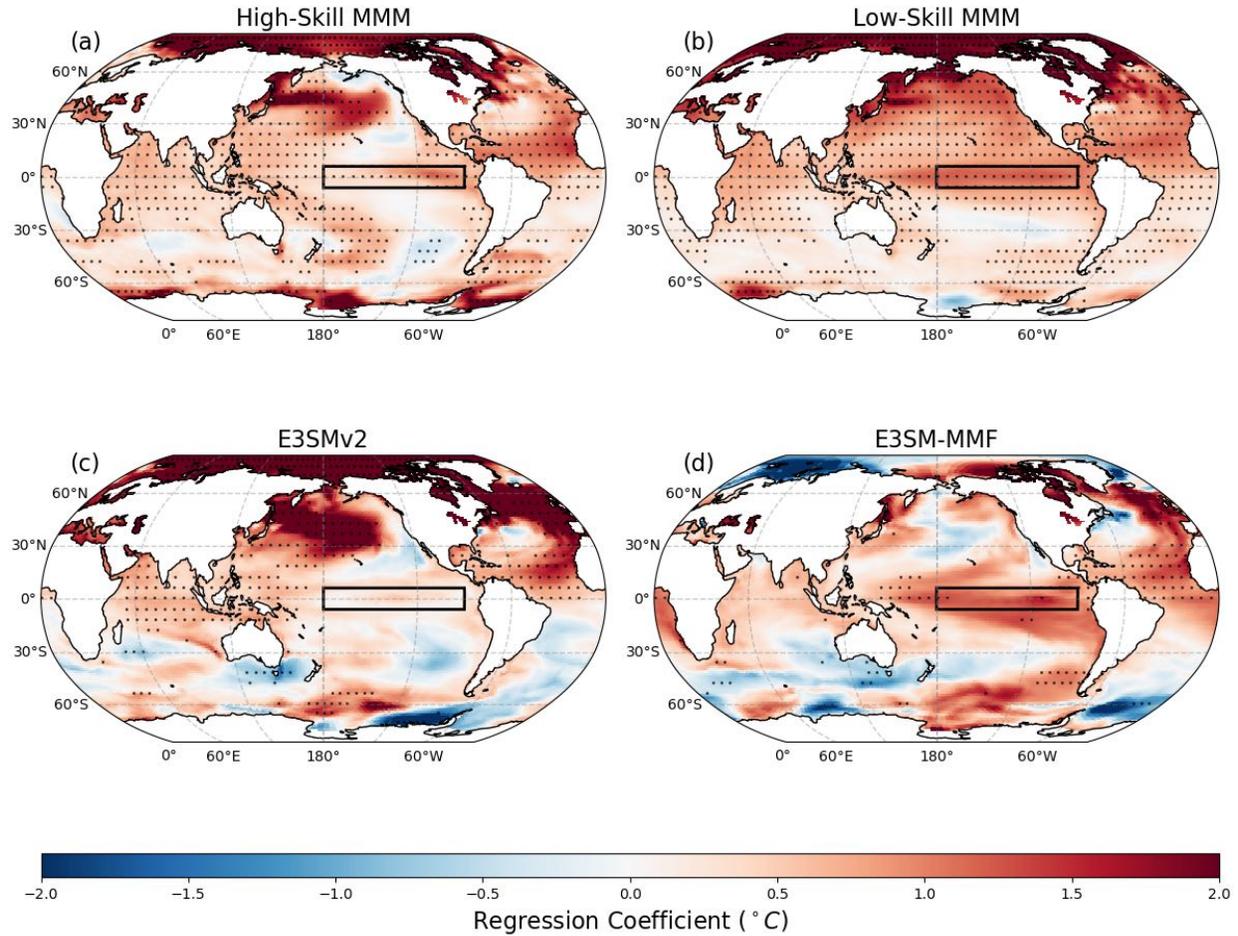

**Figure 12: Atlantic-Pacific teleconnections in E3SM configurations compared to skill groups.** Regression coefficients of SSTA onto the NTA index for (a) high-skill multi-model mean, (b) low-skill multi-model mean, (c) E3SMv2, and (d) E3SM-MMF. Statistical significance as in Figure 1.

## Insights into Model Representation of Atlantic-Pacific Interactions

The contrasting performance between E3SM configurations provides valuable insights for understanding model representation of Atlantic-Pacific interactions across CMIP6 models. Our diagnostic analysis suggests several potentially relevant physical processes that may influence how models simulate these inter-basin connections.

Precipitation-SST coupling emerges as a primary mechanism differentiating model skill in representing Atlantic-Pacific teleconnections. ERA5 shows robust precipitation responses to SST warming in both Eastern Atlantic (320°-340°E) and Western Pacific (140°-160°E). Models systematically underestimate this coupling, as evidenced in their zonal precipitation regression patterns. High-skill models generate stronger precipitation responses to NTA warming compared to low-skill models, particularly in the Atlantic sector. This enhanced precipitation response facilitates stronger ascending motion, strengthening the Atlantic-to-Pacific pathway through Walker circulation modifications.

E3SMv2 exhibits the strongest precipitation-SST coupling among high-skill models in the Atlantic region (320°E -340°E), reinforcing its Atlantic-to-Pacific teleconnection pathway (Fig. 13). In contrast, E3SM-MMF shows reversed coupling in both the tropical Atlantic (310°E-330°E) and Western Pacific (120°E-150°E), potentially indicating a dominant atmosphere-to-ocean process. During NTA warming periods, while E3SMv2 simulates enhanced ascending motion and precipitation in the Atlantic, E3SM-MMF exhibits suppressed convection that may be associated with descending motion patterns connected to Pacific circulation. In the Western Pacific region of E3SM-MMF, the observed precipitation-SST relationships appear to correspond with enhanced ascending motion in the CEP. These circulation patterns might contribute to increased Pacific influence over Atlantic processes, potentially explaining the model's tendency toward Pacific-dominated tropical teleconnections.

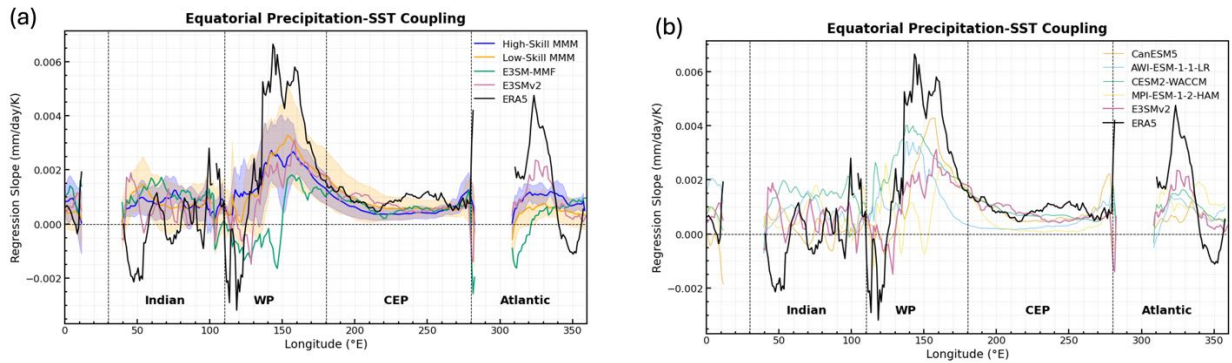

**Figure 13: Precipitation-SST coupling strength in the equatorial region.** Local precipitation-SST coupling (5°S-5°N) calculated using point-wise regression between precipitation and SST anomalies for (a) skill group comparison and (b) individual high-skill models. Positive values indicate enhanced precipitation response to local SST warming.

Our analysis suggests that low cloud-SST relationships may represent one of several relevant factors in Pacific-Atlantic interactions. Comparison with ERA5 reveals that E3SM-MMF generates significantly lower subtropical low cloud fraction during 2000-2014 in key Pacific stratocumulus regions. Specifically, the eastern (20°N-30°N, 230°E-240°E) and southern (20°S-30°S, 270°E-280°E) regions show reductions of 37.2% and 27.4% (Fig. S6), respectively. These differences in low cloud representation correspond to enhanced shortwave cloud forcing of 25.82 W/m² (37.9%) and 31.64 W/m² (44.9%) in these regions compared to Clouds and the Earth's Radiant Energy System (CERES) satellite measurement (Fig. S7), suggesting a potential warming influence on SSTs in these regions. In subtropical regions, the observed SST gradient patterns appear to align with variations in surface wind changes and latent heat flux patterns associated with WES feedback processes. Previous studies have shown that CMIP6 models with stronger low cloud feedback tend to better reproduce observed Pacific SST patterns[16]. The connection between subtropical low clouds and tropical Pacific variability operates through both local thermodynamic processes and remote wind-evaporation feedback[15]. These biases of cloud representation and radiation in the E3SM-MMF could stem from overentrainment issues common to cloud-resolving models[32] - a challenge potentially intensified by ocean-atmosphere coupling compared to atmospheric-only configurations.

**Implications for Understanding Climate Variability**

The four high-skill CMIP6 models and E3SMv2 we identified offer valuable tools for investigating decadal inter-basin interactions. Their ability to accurately reproduce the observed Atlantic influence on Pacific climate in both tropical Walker circulation and extratropical wave patterns, despite systematic westward shifts in convective centers, provides a foundation for exploring the physical mechanisms that govern large-scale climate dynamics. E3SMv2 particularly stands out by replicating the observed meridional asymmetry in Atlantic SSTA patterns, a crucial feature that remains unresolved in other high-skill models.

The limitations of observational records in studying decadal variability highlight important opportunities for controlled model experiments to advance our understanding of Atlantic-Pacific teleconnections. Pacemaker experiments that isolate local and remote forcing pathways could provide definitive tests of the causal relationships underlying the contrasting behaviors observed between high-skill and low-skill models, especially regarding how prescribed Atlantic SSTs and Pacific forcing separately influence basin interactions.

The analysis of E3SM configurations raises important questions about the relationship between model physics and teleconnection representation. While the study demonstrates that increased model complexity does not necessarily improve teleconnection representation, it stops short of fully investigating the underlying mechanisms. We encourage experimental paths that could deepen our understanding: cloud-locking experiments to quantify cloud-SST feedback effects[15], and comparative analyses of coupled versus uncoupled MMF simulations to isolate atmosphere-ocean coupling impacts[25]. These investigations would be particularly valuable in determining whether teleconnection biases stem from atmospheric processes or coupled dynamics.

# Methods

## Observations and reanalysis

For observational ground truth, we utilized two SST datasets: HadISST[33] and ERSST v5[34]. Both datasets provide monthly SST fields at 1° × 1° spatial resolution, with ERSST interpolated from its native 2° × 2° grid using bilinear interpolation. Atmospheric fields, including vertical pressure velocity (OMEGA), precipitation, and 200-hPa geopotential height (Z200), were derived from ERA5 reanalysis [35] and regridded to 1° × 1° resolution for consistency.

## Model Simulations

We analyzed historical simulations from 27 CMIP6 models and two E3SM configurations for 1950-2015[36]. These simulations follow the CMIP6 historical forcing protocol, which prescribes time-evolving greenhouse gases, anthropogenic aerosols, land use changes, and solar variability. The E3SM configurations, including the standard E3SMv2 and the experimental super-parameterized E3SM-MMF, are also driven by identical CMIP6 historical forcings to ensure consistent comparison.

The E3SM-MMF implementation represents the fully coupled historical simulation with its original configuration[19], marking a significant technical achievement that required extreme computational resources. The simulation's ocean initial condition was derived from a previous E3SMv1 simulation. Prior to the historical run, a 10-year "perpetual 1950" spin-up simulation was conducted, which showed stable global temperatures. Due to computational constraints, the E3SM-MMF configuration underwent only limited tuning compared to the comprehensive optimization process typically employed in climate models for variables such as precipitation and radiative fluxes. All models provide monthly outputs of SST, OMEGA, precipitation, and Z200.

## Analysis Methods

Our investigation of tropical Pacific-Atlantic interactions focused on two key regions. The Cold Tongue Index (CTI), defined as average detrended SST anomalies over the equatorial Pacific (180°E-280°E, 6°S-6°N), characterizes the Central-Eastern Pacific cold tongue variability. The North Tropical Atlantic (NTA) index covers 280°E-360°E, 5°N-25°N, representing tropical Atlantic Multidecadal Variability. These regions have been extensively used to study inter-basin connections on various timescales[1].

To isolate decadal variability, all variables underwent linear detrending followed by a 10-year running mean filter. This approach removes both the long-term trend and interannual variations, including ENSO signals, allowing focus on decadal-scale processes. Statistical significance is assessed using standard t-tests adapted using the effective degrees of freedom approach, accounting for temporal autocorrelation in the low pass filtered data. The degrees of freedom are calculated as N/M, where N represents the time series length and M denotes the filter width[37]. For the 10-year filtered data spanning 1950-2014, this yields approximately 6 effective degrees of freedom. A two-tailed Student's t-test determines significance at the 90% confidence level.

We developed a Total Performance Metric (TPM) combining pattern correlations between Atlantic-forced and Pacific-forced regression patterns for both tropical Walker circulation and extratropical wave responses. For each model, we computed a normalized pattern correlation score ($S_i$) as:

$$S_i = (r_i - r_{min})/(r_{max} - r_{min})$$

where $r_i$ is the correlation coefficient for model i, and $r_{min}$ and $r_{max}$ are the minimum and maximum correlations across the model ensemble. The total performance metric (TPM) combines normalized scores for Walker circulation ($S_{WC}$) and 200-hPa geopotential height ($S_{Z200}$) patterns:

$$TPM = S_{WC} + S_{Z200}$$

Models with TPM<1.0 were classified as high-skill (4 CMIP6 models plus E3SMv2), reproducing observational-like teleconnection patterns. Models with TPM>1.5 were classified as low-skill (15 models), indicating unrealistic Pacific-driven teleconnections. The model specifications and their TPM scores are provided in Supplementary Table 1.

# Acknowledgement

We gratefully acknowledge the co-funding support from the US Department of Energy (DE-SC0023368) and the National Science Foundation (NSF) Science and Technology Center (STC) Learning the Earth with Artificial Intelligence and Physics (LEAP), Award # 2019625-STC to YX and MP. WH's work on the development and application of E3SM-MMF was supported by the Exascale Computing Project (17-SC-20-SC), a joint initiative of the U.S. Department of Energy Office of Science and the National Nuclear Security Administration, as well as the Energy Exascale Earth System Model (E3SM) project, funded by the U.S. Department of Energy, Office of Science, Office of Biological and Environmental Research. This work was performed under the auspices of the U.S. Department of Energy by Lawrence Livermore National Laboratory under Contract DE-AC52-07NA27344. We thank the World Climate Research Programme for coordinating and promoting CMIP6, and the climate modeling groups for providing their model outputs through the Earth System Grid Federation (ESGF). We acknowledge the data providers of ERA5 reanalysis, HadISST, ERSST, and CERES EBAF for making their datasets publicly available. Special thanks go to Dr. Xumin Li (University of Maryland) for valuable discussions and insights, and to Yu Zhao (University of California, Irvine) for assistance with data analysis and tropical dynamics. We also appreciate the constructive comments from the anonymous reviewers that helped improve this manuscript.

# Data Availability Statement

The key datasets used in this study are publicly available through established repositories. The CMIP6 model outputs are accessible through the Earth System Grid Federation (ESGF) nodes (https://esgf-node.llnl.gov/projects/cmip6/). The ERA5 reanalysis data were obtained from the Copernicus Climate Data Store (https://cds.climate.copernicus.eu). SST observations from HadISST and ERSST datasets are available from the Met Office Hadley Centre (https://www.metoffice.gov.uk/hadobs/hadisst/) and NOAA's National Centers for Environmental Information (https://www.ncei.noaa.gov/products/extended-reconstructed-sst), respectively. The CERES radiation data were accessed through NASA's CERES data portal (https://ceres.larc.nasa.gov/data/). The preprocessed data used in this study, including E3SMv2 and E3SM-MMF models, are available through Zenodo at https://doi.org/10.5281/zenodo.14884585. The analysis scripts used to process the data and generate the results will be made available through GitHub at https://github.com/SciPritchardLab/Tropical-Atlantic-Pacific-Multi-decadal-Teleconnections.

# SI- Diagnosing Biases in Tropical Atlantic-Pacific Multi-Decadal Teleconnections Across CMIP6 and E3SM Models

**Yan Xia[1]\*, Yong-Fu Lin[1], Jin-Yi Yu[1], Walter Hannah[2], Mike Pritchard[1]**

[1]Department of Earth System Science, University of California, Irvine, California, USA

[2]Lawrence Livermore National Laboratory, Livermore, CA

Corresponding author: Yan Xia (yxia24@uci.edu)

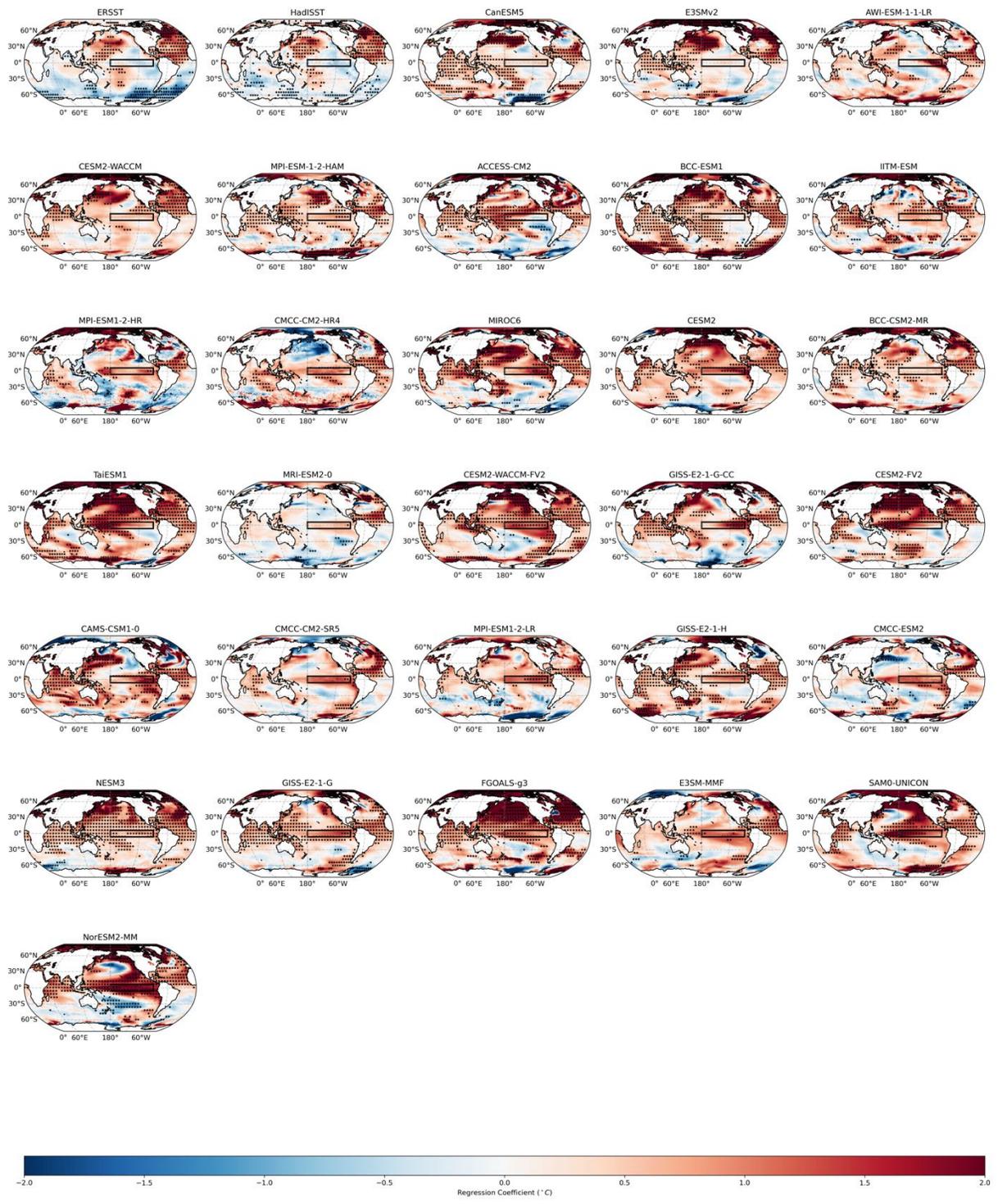

**Figure S1.** Same as Figure 1 but for individual observation/model regression of low pass filtered global SSTA onto the NTA index during 1950-2014. The first two panels show observations. The models are arranged by TPM ranking from lowest (best performance) to highest (worst performance), with five high-skill models (TPM < 1) following ERA5, while the last fifteen models show low skill (TPM > 1.5).

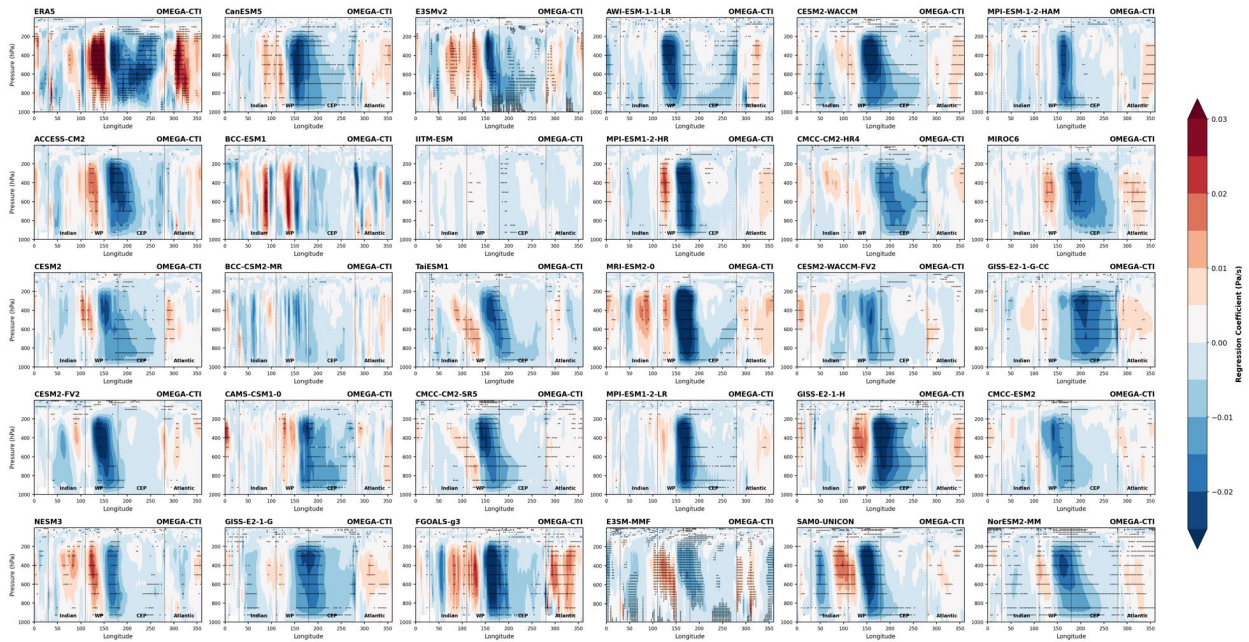

**Figure S2.** Same as Figure 2 but for individual regression of low pass filtered Walker circulation anomalies onto the CTI index during 1950-2014. The first panel shows ERA5 reanalysis data. The models are arranged by TPM ranking from lowest (best performance) to highest (worst performance), with five high-skill models (TPM < 1) following ERA5, while the last fifteen models show low skill (TPM > 1.5).

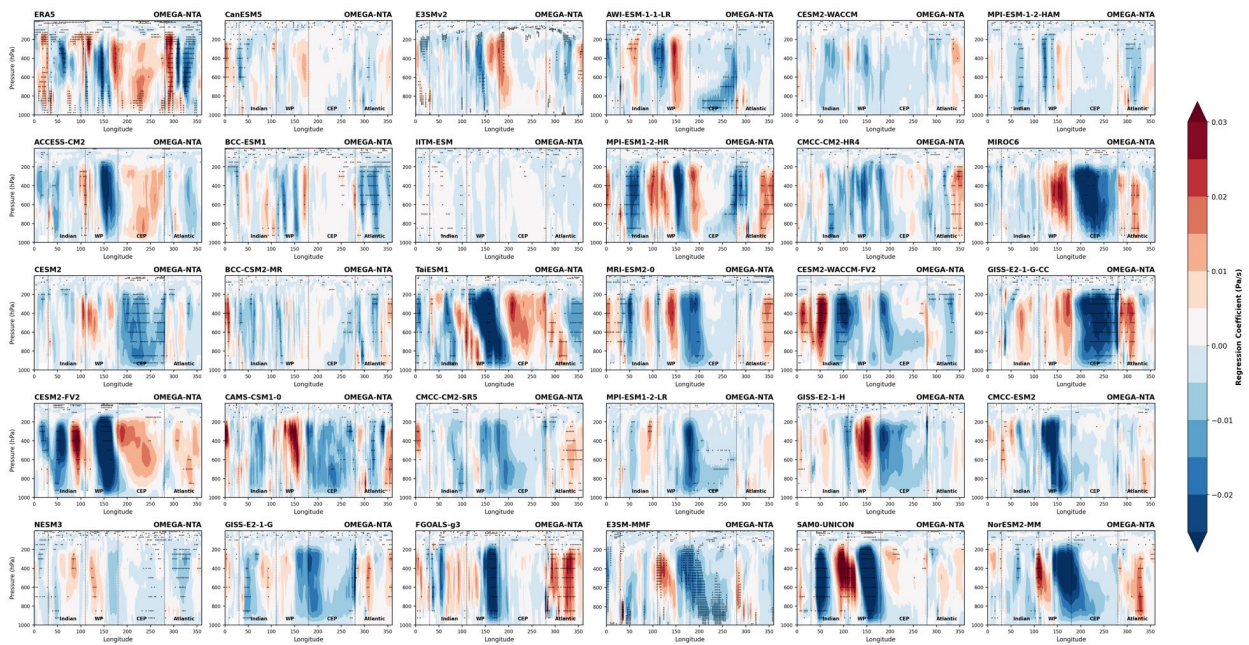

**Figure S3.** Same as Figure S2 but for individual regression of low pass filtered Walker circulation anomalies onto the NTA index during 1950-2014.

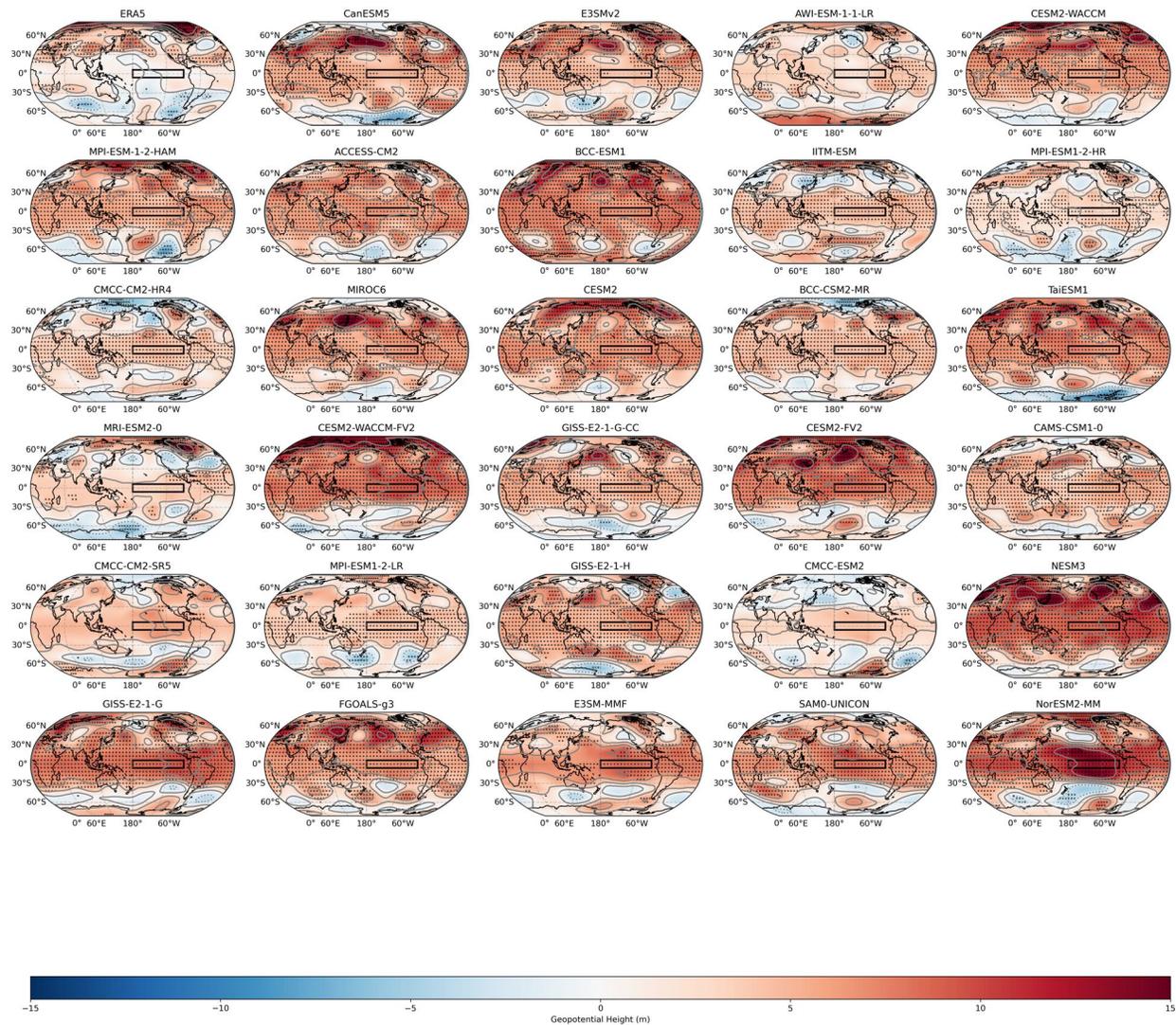

**Figure S4.** Same as Figure 3 but for individual regression of low pass filtered Z200 anomalies onto the NTA index during 1950-2014. The first panel shows ERA5 reanalysis data. The models are arranged by TPM ranking from lowest (best performance) to highest (worst performance), with five high-skill models (TPM < 1) following ERA5, while the last fifteen models show low skill (TPM > 1.5).

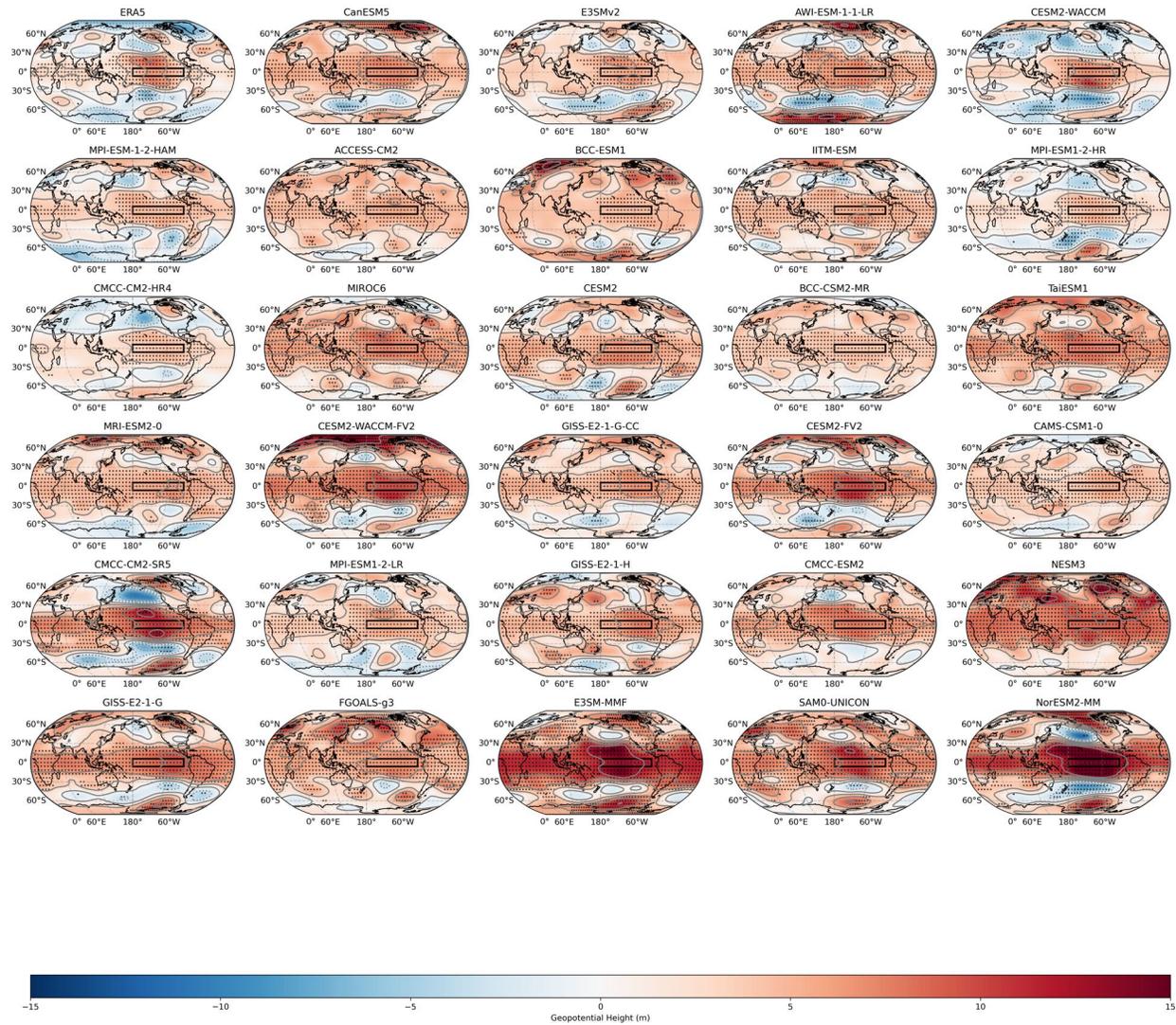

**Figure S5.** Same as Figure S4 but for individual regression of low pass filtered Z200 anomalies onto the CTI index during 1950-2014.

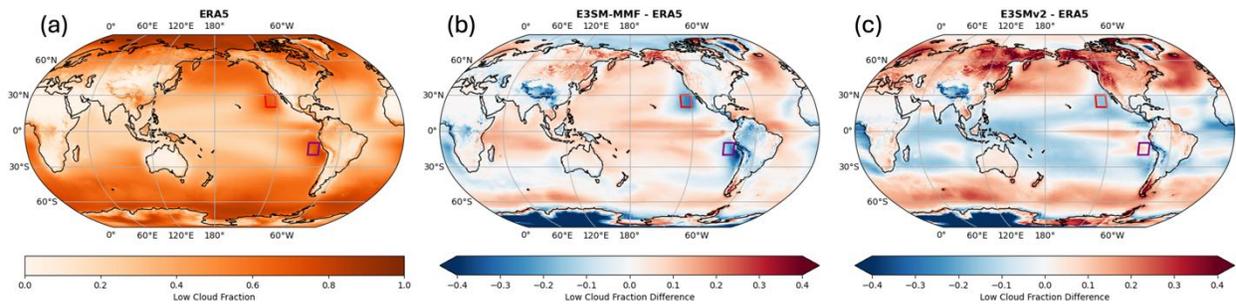

**Figure S6.** Mean global distribution of low cloud fraction during 2000-2014. (a) ERA5 reanalysis baseline (0-1.0). (b) E3SM-MMF minus ERA5 difference. (c) E3SMv2 minus ERA5 difference. Red and purple boxes denote eastern and southern Pacific stratocumulus regions, respectively. Differences range from -0.4 to 0.4.

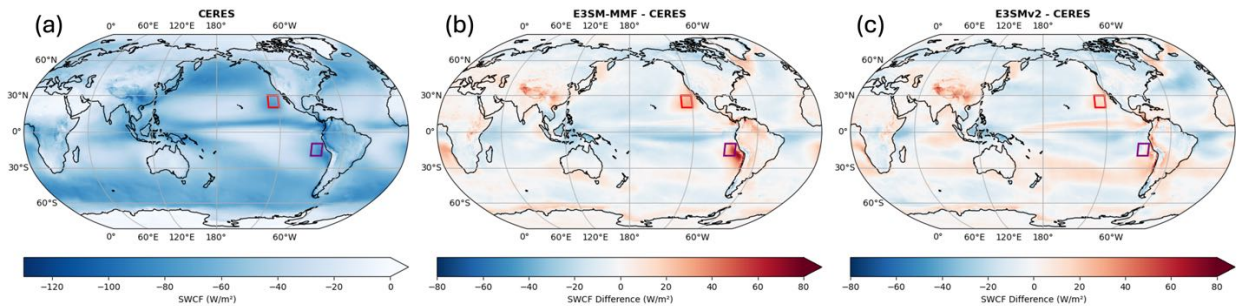

**Figure S7.** Mean global distribution of shortwave cloud radiative forcing (SWCF) during 2000-2014. (a) CERES observations (W/m²). (b) E3SM-MMF minus CERES difference. (c) E3SMv2 minus CERES difference. Red and purple boxes denote eastern and southern Pacific stratocumulus regions, respectively. Differences range from -80 to 80 W/m².

**Table S1.** List of CMIP6 models and E3SM configurations analyzed in this study, ranked by their Total Performance Metric (TPM) values. As will be described in Section 3.2, TPM quantifies model fidelity in representing decadal Atlantic-Pacific teleconnections, with lower values indicating better agreement with observations.

| Ranking | Models | TPM |
| --- | --- | --- |
| 1 | CanESM5 | 0.03 |
| 2 | E3SMv2 | 0.23 |
| 3 | AWI-ESM-1-1-LR | 0.79 |
| 4 | CESM2-WACCM | 0.92 |
| 5 | MPI-ESM-1-2-HAM | 0.96 |
| 6 | ACCESS-CM2 | 1.24 |
| 7 | BCC-ESM1 | 1.26 |
| 8 | IITM-ESM | 1.28 |
| 9 | MPI-ESM1-2-HR | 1.33 |
| 10 | CMCC-CM2-HR4 | 1.34 |
| 11 | MIROC6 | 1.35 |
| 12 | CESM2 | 1.37 |
| 13 | BCC-CSM2-MR | 1.39 |
| 14 | TaiESM1 | 1.52 |
| 15 | MRI-ESM2-0 | 1.52 |
| 16 | CESM2-WACCM-FV2 | 1.53 |
| 17 | GISS-E2-1-G-CC | 1.54 |
| 18 | CESM2-FV2 | 1.54 |
| 19 | CAMS-CSM1-0 | 1.60 |
| 20 | CMCC-CM2-SR5 | 1.62 |
| 21 | MPI-ESM1-2-LR | 1.65 |
| 22 | GISS-E2-1-H | 1.70 |
| 23 | CMCC-ESM2 | 1.78 |
| 24 | NESM3 | 1.87 |
| 25 | GISS-E2-1-G | 1.90 |
| 26 | FGOALS-g3 | 1.92 |
| 27 | E3SM-MMF | 1.92 |
| 28 | SAM0-UNICON | 1.99 |
| 29 | NorESM2-MM | 2.00 |